\documentclass[screen,acmsmall]{acmart}

\AtBeginDocument{%
  }

\setcopyright{acmlicensed}
\copyrightyear{2025}
\acmYear{2025}
\acmDOI{XXXXXXX.XXXXXXX}
\acmISBN{978-1-4503-XXXX-X/2018/06}

\usepackage{pdflscape}
\usepackage{tabularx}
\usepackage{multirow}
\usepackage{array}
\usepackage{tikz}
\usetikzlibrary{shapes, positioning, arrows.meta}
\usepackage{booktabs}  % optional, for prettier lines
\usepackage{ragged2e} 
\usepackage{subcaption}
\usepackage{hyperref} 
\usepackage{titlesec} % for \subsubsubsection

\newcommand{\xhdr}[1]{\vspace{1.5mm}\noindent\textbf{#1}}
\begin{document}

\title{Understanding Community-Level Blocklists in Decentralized Social Media}

\author{Owen Xingjian Zhang}
\email{owenz@princeton.edu}
\orcid{0009-0008-2949-7379}
\affiliation{%
  \institution{Princeton University}
  \country{USA}
}

\author{Sohyeon Hwang}
\email{sohyeon@princeton.edu}
\orcid{0000-0001-8415-7395}
\affiliation{%
  \institution{Princeton University}
  \country{USA}
}

\author{Yuhan Liu}
\email{yl8744@princeton.edu}
\orcid{0000-0001-6852-6218}
\affiliation{%
  \institution{Princeton University}
  \country{USA}
}

\author{Manoel Horta Ribeiro}
\email{manoel@cs.princeton.edu}
\orcid{0000-0002-6159-9657}
\affiliation{%
  \institution{Princeton University}
  \country{USA}
}

\author{Andrés Monroy-Hernández}
\email{andresmh@princeton.edu}
\orcid{0000-0003-4889-9484}
\affiliation{%
  \institution{Princeton University}
  \country{USA}
}

\renewcommand{\shortauthors}{Zhang et al.}

\begin{abstract}
Community-level blocklists are key to content moderation practices in decentralized social media. These blocklists enable moderators to prevent other communities, such as those acting in bad faith, from interacting with their own --- and, if shared publicly, warn others about communities worth blocking. Prior work has examined blocklists in centralized social media, noting their potential for collective moderation outcomes, but has focused on blocklists as individual-level tools. 
To understand how moderators perceive and utilize community-level blocklists and what additional support they may need, we examine social media communities running Mastodon, an open-source microblogging software built on the ActivityPub protocol. 
We conducted (1) content analysis of the community-level blocklist ecosystem, and (2) semi-structured interviews with twelve Mastodon moderators.  
Our content analysis revealed wide variation in blocklist goals, inclusion criteria, and transparency. Interviews showed moderators balance proactive safety, reactive practices, and caution around false positives when using blocklists for moderation. They noted challenges and limitations in current blocklist use, suggesting design improvements like comment receipts, category filters, and collaborative voting. 
We discuss implications for decentralized content moderation, highlighting trade-offs between openness, safety, and nuance; the complexity of moderator roles; and opportunities for future design.

\end{abstract}




\maketitle

\section{Introduction}
\label{sec:intro}
In \textit{The Open Society and Its Enemies}~\citep{popper2012open}, philosopher Karl Popper introduces the ``paradox of tolerance'': unlimited tolerance can undermine itself by allowing intolerance to flourish.
This philosophical insight highlights communities' ongoing challenge: \textit{determining which behaviors to accept and which to reject to maintain inclusive and civil discourse.}

Online communities operationalize this principle through content moderation, \textit{the process of shaping information exchange and user activity by deciding upon and filtering posted content according to established rules} \citep{kiesler2012regulating,zeng2022content, grimmelmann2015virtues}. 
Previous studies have extensively examined content moderation practices at three levels in centralized social media: platform, community, and personal levels.
At the (centralized) platform level, moderation typically involves a combination of algorithmic systems and human moderators --- either paid or volunteer --- who enforce site-wide policies and manage user-generated content \citep{gillespie2018custodians, roberts2019behind, chandrasekharan2017you}.
At the community level, collective moderation approaches --- such as automated moderation bots and shared community norms --- are meant to empower groups to collectively regulate their own spaces \citep{jhaver2019human, hwang2024adopting, seering2020reconsidering}. 
At the personal level, moderation actions such as reporting, flagging, and personal blocklists aim to help users mitigate harassment and unwanted interactions \citep{crawford2016flag, jhaver2018blocklists}. 

However, these insights predominantly reflect centralized contexts, leaving significant gaps in understanding moderation within decentralized platforms, where even the very notion of a ``platform'' is contested as systems are built based on ``protocols'' that define the building blocks of digital infrastructure~\cite{masnick2019}. 
Consider, for example, Mastodon~\cite{mastodon}, a widely used open-source software implementation of the ActivityPub protocol~\cite{activitypub2018}. Mastodon is a decentralized alternative to mainstream centralized social media platforms like X. With Mastodon built for the ActivityPub protocol, people can create or join independently operated servers commonly referred to as ``\textit{instances}'' (e.g., \texttt{mastodon.social}, \texttt{mstdn.party}), which are part of the ``\textit{Fediverse}.'' The Fediverse refers to the network of thousands of such instances running Mastodon and other ActivityPub implementations. Instances can interact with one another through the ActivityPub protocol. However, each instance in the Fediverse is independently administered by a user or group who sets local rules and moderation policies, effectively functioning as a self-governed online community. This independence allows communities to reflect local values but also creates challenges around consistency and coordination across instances~\citep{datta2010decentralized, rozenshtein2023fediverse, hassan2021impact, anaobi2023will, hwang2025trust}.

Importantly, instances have limited capacity to moderate beyond their own community; instance A has no say over what is or is not allowed to be posted in instance B, even if they are connected. While more sophisticated content moderation mechanisms are unavailable, the key moderation mechanism available to Mastodon instances to handle cross-instance interactions is \textit{community-level blocklists}. They allow users in roles of community moderators to protect their communities by blocking entire instances they believe are problematic~\cite{bono2024mastodon, colglazier2024effects}.
These decisions can potentially shape a user's experience in Mastodon completely. For example, if the instance users reside in blocks or are blocked by the most popular instances, they would be cut from a broad swathe of content and interactions in the Fediverse. As social media systems like Mastodon become more widely adopted \citep{zhang2024trouble, liu2025understanding}, understanding blocklists in this context is crucial for improving content moderation in decentralized settings. 
However, despite their prominence, little is known about how Mastodon moderators currently manage these blocklists, what tools they use, and what additional support they need.  

\xhdr{Present Work.}
To address the current lack of understanding surrounding community-level blocklists, this study adopts a mixed-methods approach guided by three research questions:

\begin{itemize}
\item \textbf{RQ1:} What is the current landscape of community-level blocklists?
\item \textbf{RQ2:} How do moderators interpret and apply community-level blocklists in practice?
\item \textbf{RQ3:} What design improvements could enhance the usability and effectiveness of tools for managing community-level blocklists?
\end{itemize}

\noindent
We investigated these questions through content analysis and semi-structured interviews.
First, we characterized the usage of blocklists and the capabilities of blocklist-related tools (\textbf{RQ1}). Second, we conducted semi-structured interviews with 12 Mastodon moderators to explore their experiences, practices, and perceptions of blocklists (\textbf{RQ2}). Finally, we developed and presented a prototype blocklist management interface to these moderators, using it as a design probe to identify needs and elicit feedback on potential future design work (\textbf{RQ3}).

\xhdr{Summary of Results.}
We found that existing blocklists and tools differ substantially in their purposes, inclusion criteria, transparency, and accountability. Some prioritize broad consensus across instances, while others focus on severity-based or manual curation  (\textbf{RQ1}). Through interviews, we identified how moderation styles shape the understanding and usage of blocklists. Different styles embody different trade-offs of how open, safe, and context-specific moderation should be (\textbf{RQ2}). 
Finally, we find several venues for improvement in the design of blocklist tools, including category filters, severity toggles, comment receipts, and collaborative moderation functionalities (\textbf{RQ3}). 

\xhdr{Implications.}
Our work characterizes the current state of community-level blocklists and related tools on a growing decentralized social media platform. We expect our results to lay the foundation for future work experimenting with how different moderation practices and additional design features may help support more flexible and efficient blocklist-based content moderation.

\section{Related Work}

\subsection{Content Moderation in Centralized Social Media}

Content moderation refers to the governance of digital interactions by approving, filtering, or removing content in accordance with platform rules or community norms \citep{kiesler2012regulating,zeng2022content,grimmelmann2015virtues}.
Platforms--digital services that mediate user interaction and content exchange--have often drawn the most attention in discussions of moderation~\citep{gillespie2010politics,gorwa2019platform}.
Yet a growing body of research shows that individuals and communities also actively shape their online environments through moderation practices \citep{jhaver2023personalizing,seering2020reconsidering,schneider2022implicit,kharazian2023governance,chandrasekharan2018internet}. Below, we briefly discuss past work centered around platform, personal, and community moderation.

\textit{Platform moderation} refers to actions taken by platform operators --- typically corporations --- through algorithmic filters or hired human moderators. Within platforms like Facebook, Amazon, or Tinder, automated systems are widely used to enforce site-wide policies and manage large volumes of content~\citep{jhaver2023personalizing, roberts2019behind, duffy2023platform, douek2022siren, gillespie2020custodians, jhaver2021evaluating, pasquale2015black}. 
Past research has critically examined these algorithmic systems, noting their effectiveness and necessity~\citep{gillespie2018scale}, but also frequent transparency issues, exacerbating accountability and trust concerns \citep{gillespie2018custodians,crawford2016flag, dias2021fighting}. In parallel, platforms deploy paid human moderators to handle content that algorithms cannot adequately address \citep{roberts2019behind, mcgillicuddy2016controlling}. In that context, past work has raised ethical concerns around exploitation, psychological harm, and poor working conditions \citep{roberts2019behind}. 
Finally, platforms often take centralized, targeted decisions to ban controversial influencers or communities. Recent work has suggested that these bans can effectively reduce the reach and toxicity of said entities within the platforms~\citep{chandrasekharan2017you, horta2021platform}.

\textit{Personal moderation} refers to actions taken by individual users to manage their online experiences, including flagging, blocking, and using platform-provided tools \citep{kalch2017replying, kraut2012building, mitchell2005protecting}. These tools are often limited, offering little transparency into how reports are handled or why moderation decisions are made \citep{crawford2016flag, myers2018censored}. In response, users have developed and shared their own moderation resources \citep{kiene2019technological, geiger2013levee}, such as crowd-sourced blocklists that resist dominant norms and foster counterpublics \citep{geiger2016blockbots, jhaver2018blocklists}. Research in social computing has also introduced platform-independent tools like Squadbox for mitigating harassment \citep{mahar2018squadbox}, and customizable filters that support proactive, personalized moderation~\citep{huang2024opportunities, jhaver2022designing}.

\textit{Community moderation}, the focus of this work, refers to collective decision-making within user groups to govern behavior in shared virtual spaces, as seen on platforms like Reddit, Discord, and Twitch.  
Research has examined how communities establish norms \citep{seering2017shaping, black2011self, joyce2013rules, matias2018civilservant, weld2022what}, enforce them through moderation tools \citep{jhaver2019human, hwang2024adopting, li2022all, jiang2019moderation}, and navigate the emotional and material labor involved \citep{seering2022metaphors, schopke2022why, dosono2019moderation}. 
Previous work has also specifically examined how moderators across various platforms conceptualize their roles \citep{seering2022metaphors}, provide substantial monetary value \citep{li2022all} and emotional labor \citep{schopke2022why, dosono2019moderation} to sustain safe spaces online, and can subvert platform power through collective action \citep{matias2016goingdark}. Across these examples, community moderation emerges as a collective effort to exercise self-governance and maintain community-defined standards, and at the same time, to adapt to platform-level constraints such as limited tooling \citep{kiene2019technological} or ambiguous rules \citep{matei2011wikipedia}.

\subsection{The Unique Challenges of Decentralized Social Media}

Decentralized social media \citep{thompson2022web3} has gained traction in response to growing public scrutiny of harassment, ethical labor, and algorithmic bias on their centralized corollaries \citep{massanari2017gamergate, li2022ethical, li2023dimensions, duffy2023platform, caplan2020tiered}.
They are built on open protocols such as ActivityPub \citep{activitypub2018}, which define standard methods for servers to exchange messages, follow users, and share posts. Through software that implements this protocol, like Mastodon, anyone can create their own communities by setting up a server (called an ``instance'') that can interact with others but remain independently governed  \citep{masnick2019}. 

Unlike centralized systems, no single entity governs the network of instances (colloquially called the ``Fediverse''); instead, each community sets its own rules, moderation policies, and membership criteria based on its values and context \citep{anaobi2023will, hassan2021pleroma, hassan2021impact, nicholson2023mastodon}. 
As such, community moderation plays a significant role in these systems compared to centralized contexts, where platform operators act as a primary enforcement authority that can limit the agency of individual users and communities. Emerging scholarship suggests that moderation models from centralized systems do not easily translate to decentralized environments, given the fundamentally different technical architectures and governance structures \citep{abbing2023decentralised, guidi2018managing, zulli2023digital, tosch2024privacy}.

Recent work focused on decentralized social media in particular has examined the experiences of community admins, highlighting both familiar patterns and novel challenges unique to decentralized contexts \citep{zhang2024trouble,hassan2021pleroma,rozenshtein2023fediverse,anaobi2024improving,anaobi2023will,spencersmith2025labourpains}. 
A key challenge is that in decentralized platforms, content flows across communities with differing values and goals, increasing the likelihood of conflicts due to incompatible community norms \citep{hwang2025trust, kumar2018community, zhang2024debate}. While differences in community norms are common~\citep{chandrasekharan2018internet, fiesler2018reddit, weld2022what}, the interconnected yet independent structure of decentralized platforms puts conflicting norms in contact with one another, such that their incompatibilities are salient and must be resolved \citep{hwang2025trust}. At the same time, resolving conflicts between communities appears to be particularly challenging, as decentralized platforms lack robust moderation infrastructure and tools for transparency and accountability, compounding difficulties in cultivating user trust and understanding across diverse instances \citep{roth2024federated}. 

Admins on decentralized social media routinely moderate by removing harmful content, issuing warnings, and blocking problematic users or instances \citep{zhang2024trouble,hassan2021pleroma}, a process that demands significant labor and poses risks of burnout similar to those seen on centralized platforms \citep{anaobi2023will, schopke2022why}. In response, organizers, developers, and designers have begun developing tools and practices to ease these burdens. One prominent approach is using community-level blocklists: lists of instances to block that can be shared across communities. These blocklists and related management tools offer a way to streamline moderation by providing reusable resources for identifying harmful actors. Today, they represent the primary mechanism for addressing inter-community conflicts on decentralized platforms. Accordingly, this work examines blocklists as a central tool for community moderation in decentralized social media.

\subsection{Blocklists as a Community-Level Tool}

Blocklists have mainly been studied as tools for individual-level moderation on centralized platforms, where users block other users to manage harassment \citep{geiger2016blockbots,jhaver2018blocklists}. 
When widely shared and adopted, these lists can enable collective bottom-up resistance and form protective counterpublics~\citep{geiger2016blockbots}, especially for marginalized groups at higher risk of online harm \citep{blackwell2017classification, jhaver2018blocklists}. 
On decentralized social media, blocklists provide a similar individual-level function but operate at the community level, blocking instances rather than users.

Community-level blocklists are a core moderation strategy in Mastodon. Administrators can self-host instances, block others entirely, and optionally share their own blocklists. Many moderators also rely on these shared blocklists to shield their users from spam, harassment, and other harmful content. These are distributed via public repositories and maintained by individuals, teams, or trusted groups. Several prominent shared blocklists have emerged across the Fediverse, accompanied by tools that support their integration and management.\footnote{\url{https://jaygraber.medium.com/designing-decentralized-moderation-a76430a8eab}}\textsuperscript{,}\footnote{\url{https://seirdy.one/posts/2023/05/02/fediverse-blocklists/}}\textsuperscript{,}\footnote{\url{https://writer.oliphant.social/oliphant/blocklists}} Prior work highlights how these tools enable a shift from individual to collective approaches in content moderation \citep{bono2024mastodon,hwang2025trust,zhang2024trouble}, allowing communities to coordinate moderation at scale.

However, despite their growing adoption, community-level blocklists face unresolved concerns. Their curation processes are often opaque, making it difficult to assess how decisions are made and by whom \citep{feal2021blocklist}. This lack of transparency can complicate trust and limit uptake, particularly in communities prioritizing mutual aid and accountability \citep{melder2025blocklist}. Moreover, blocklists may reproduce well-known issues from individual-level moderation, including over-blocking and conflicting definitions of harm \citep{jhaver2018blocklists}. As such, scholars have called for more comprehensive, mixed-method research into blocklists' structure, governance, and impacts, along with new designs to support more transparent, collaborative moderation practices \citep{melder2025blocklist, gillespie2020expanding}.

Given these unresolved tensions and the gaps identified in prior work, there is a pressing need to better understand how community-level blocklists are used and perceived in practice. Examining how moderators adopt, adapt, and make sense of these tools is essential for clarifying their role within decentralized moderation ecosystems. Such insights are foundational for informing the design of decentralized social media's more transparent, accountable, and community-aligned moderation infrastructures.

\section{Methods}
\subsection{Empirical Setting}

To study community-level blocklists in decentralized social media, we examine instances on the Fediverse, which we briefly introduced in Section \ref{sec:intro}, focusing on Mastodon instances as our empirical setting of interest. Mastodon is a widely used open-source software enabling individuals to run their own instances on the Fediverse. Throughout this work, we define \textbf{moderators} as individuals responsible for managing and moderating interactions on their respective Mastodon instances.%
\footnote{We use ``moderators'' to collectively refer to various Mastodon moderation roles (Owner, Administrator, Moderator).}

Like many other decentralized social media platforms, Mastodon supports \textbf{community-level blocklists}—lists that allow moderators to block entire instances from interacting with their community. Moderators may choose to keep their blocklists private or share them publicly as references for others curating their own lists. Over time, some of these publicly shared blocklists—referred to as \textbf{shared community-level blocklists}—have gained widespread adoption across Mastodon communities. This work presents an overview of community-level blocklists, several popular shared community-level blocklists, and third-party tools designed to support their creation and management (\textbf{blocklist-related tools}).

\subsection{Content Analysis}
\label{sec:content_analysis}

To answer \textbf{RQ1}, we analyzed an overview of community-level blocklists, widely shared community-level blocklists, and related tools used by Mastodon moderators.

\subsubsection{Overview of Community-Level Blocklists.}

We collected community-level blocklist data from Mastodon instances with at least 10 active users using the \texttt{instances.social} API. Then, we employed automated scraping\footnote{We used a web scraping tool called Octoparse. Read more at https://www.octoparse.com/} to extract data from each instance’s publicly accessible ``About'' page, specifically retrieving listed blocked domains and blocking reasons (when available). We systematically parsed these textual rationales and manually categorized them into thematic groups (e.g., spam, harassment, hate speech). Additionally, we documented instance size (number of active users) to analyze patterns in blocklist sharing.

\subsubsection{Shared Community-Level Blocklists.}

Drawing on community-curated lists,\footnote{\url{https://codeberg.org/nev/awesome-fediadmin/\#basic-blocklists}}\textsuperscript{,}\footnote{\url{https://connect.iftas.org/library/iftas-documentation/cariad-policy/}}\textsuperscript{,}\footnote{\url{https://github.com/ineffyble/mastodon-block-tools}}\textsuperscript{,}\footnote{\url{https://writer.oliphant.social/oliphant/the-oliphant-social-blocklist}} we selected five blocklists (Seirdy Tier-0, FediNuke, Garden Fence, CARIAD, and IFTAS-DNI). Our selection criteria prioritized public accessibility, community-driven maintenance, extensive documentation, and widespread adoption within the Mastodon community. To analyze and characterize current blocklists, the first author thoroughly read the page and documentation of every blocklist in the set. They then created a rough list of dimensions of key blocklist information that would serve as categories to code each blocklist. After getting iterative feedback from other authors, the first author coded each blocklist along the categories, which led to minor adjustments of the categories to ensure conceptual consistency and alignment.

\subsubsection{Blocklist-Related Tools.}
We also selected four blocklist-related tools (FediCheck, FediBlockHole, Fediseer, and The Bad Space) from the same set of community-curated resources. Our selection criteria prioritized public accessibility, extensive documentation, and support for different stages of the moderation process, including discovery, evaluation, and implementation. The categorization process was repeated separately for blocklist-related tools, resulting in a distinct set of categories tailored to how each tool supports moderation. The final categories and their definitions are shown in Table~\ref{tab:combined-categories}. The results are shown in Table~\ref{tab:blocklists-summary} and Table~\ref{tab:tools-summary}.

\renewcommand{\tabularxcolumn}[1]{m{#1}}
\newcolumntype{C}[1]{>{\centering\arraybackslash}m{#1}}
\renewcommand{\arraystretch}{1.1} % single-line spacing

\begin{table}[t]
\small
\centering
\caption{Combined analytical categories for Blocklists and Blocklist-related Tools}
\label{tab:combined-categories}
\begin{tabularx}{\textwidth}{|C{1.2cm}|m{2.5cm}|X|}
\hline
\textbf{Type} & \textbf{Category} & \textbf{Definition} \\
\hline
\multirow{7}{*}{\centering Blocklist} & Purpose & Initial motivation and ultimate goal behind creating the blocklist. \\
\cline{2-3}
 & Criteria & Specific requirements determining instance inclusion or exclusion. \\
\cline{2-3}
 & Transparency & Explanation of review processes, resources used, and how these resources informed blocklisting decisions. \\
\cline{2-3}
 & Distribution & Public availability and frequency of updates. \\
\cline{2-3}
 & Limitations & Limitations or challenges explicitly reported by the blocklist creators. \\
\cline{2-3}
 & Other Resources & Additional recommended blocklists or tools for moderators beyond this blocklist. \\
\cline{2-3}
 & Additional Content & Supplementary contributions or content provided by blocklist maintainers are beneficial for moderators. \\
\hline
\multirow{5}{*}{\centering Tool} & Summary & Brief overview describing the main functionality of the tool. \\
\cline{2-3}
 & Purpose & Motivation and objectives behind the creation of the tool. \\
\cline{2-3}
 & Technical Access & Means through which users interact with or technically access the tool. \\
\cline{2-3}
 & Decision Type & Methodology or basis by which the tool makes or facilitates moderation decisions. \\
\cline{2-3}
 & Tags/Categories & System for tagging or categorizing moderation entries provided by the tool. \\
\hline
\end{tabularx}

\end{table}

\subsection{Interviews}
To answer \textbf{RQ2} and \textbf{RQ3}, we conducted semi-structured interviews with 12 Mastodon moderators, focusing on how they perceive and utilize blocklists as well as blocklist-related tools.

\subsubsection{Participant Recruitment. }
To recruit participants, we sent direct invitations via Mastodon direct messages (DMs) and emails to Mastodon instance owners identified through filtering on \texttt{instances.social }for instances with more than 10 users. Additionally, we employed snowball sampling by inviting recommendations from initial respondents to expand our participant pool. Participants interested in the preliminary demographic survey were then contacted to schedule interviews.

\subsubsection{Study Protocol. }
\begin{table}[t]
\footnotesize
\vspace{2mm}
\caption{Participant demographics and instance details.}
\label{tab:participant_pool}
\begin{tabular}{llllll}\toprule
ID & Role & Country & Gender & Instance Registration & Active Users \\
\midrule
P1 & Owner & France & Male & Open (by admin approval) & 300+ \\
P2 & Owner & USA & Female & Closed & 100+ \\
P3* & Owner & New Zealand & Male & Open, Open (by admin approval) & <100 (3 instances) \\
P4 & Moderator & USA & Prefer not to say & Open (by admin approval) & 300+ \\
P5 & Administrator & USA & Non-binary & Open (by admin approval) & 700+ \\
P6* & Owner & Netherlands & Male & Open & 200+, 9700+ \\
P7 & Owner & Canada & None & Open (by admin approval) & 500+ \\
P8 & Owner & USA & Male & Open & 100+ \\
P9 & User & USA & Non-binary & Open (by admin approval) & 100+ \\
P10 & Owner & China & Prefer not to say & Open (by admin approval) & 600+ \\
P11* & Owner & Canada & Female & Closed, Open (by admin approval) & 100+, 800+ \\
P12 & Owner & South Korea & Non-binary & Open & 100+ \\
\bottomrule
\vspace{1mm}

\end{tabular}
\raggedright \small \textit{Note:} Participants marked with (*) moderated multiple instances. ``Moderator'' includes Owner, Administrator, and Moderator roles, as all have moderation access. P9 was included as an experienced Fediverse writer and is an exception.

\end{table}

We began the study with a pre-interview demographic survey and informed consent procedures. We conducted semi-structured interviews over Zoom with 12 Mastodon moderators, aged between 26 and 55, from seven countries (USA, France, New Zealand, Netherlands, Canada, China, and South Korea). Participants included six men, two women, two non-binary individuals, and two who preferred not to specify their gender (see Table~\ref{tab:participant_pool} for details).

Each interview lasted approximately one hour and was structured into two segments. The first 30 minutes explored participants' current practices in using and managing blocklists (RQ2). The second half introduced a prototype moderation tool, a \textbf{demo system} (\S \ref{sub:demo_system}) we developed to support community-level blocklist management. Using the system as a probe, we asked participants to evaluate its key features and share their preferences for additional functionalities (RQ3). We recorded the interviews with informed consent, and compensated participants for their time with a \$15 payment via Zelle. Our university's Institutional Review Board approved this work.

\begin{figure}[t]
    \centering
    \includegraphics[width=1.0\linewidth]{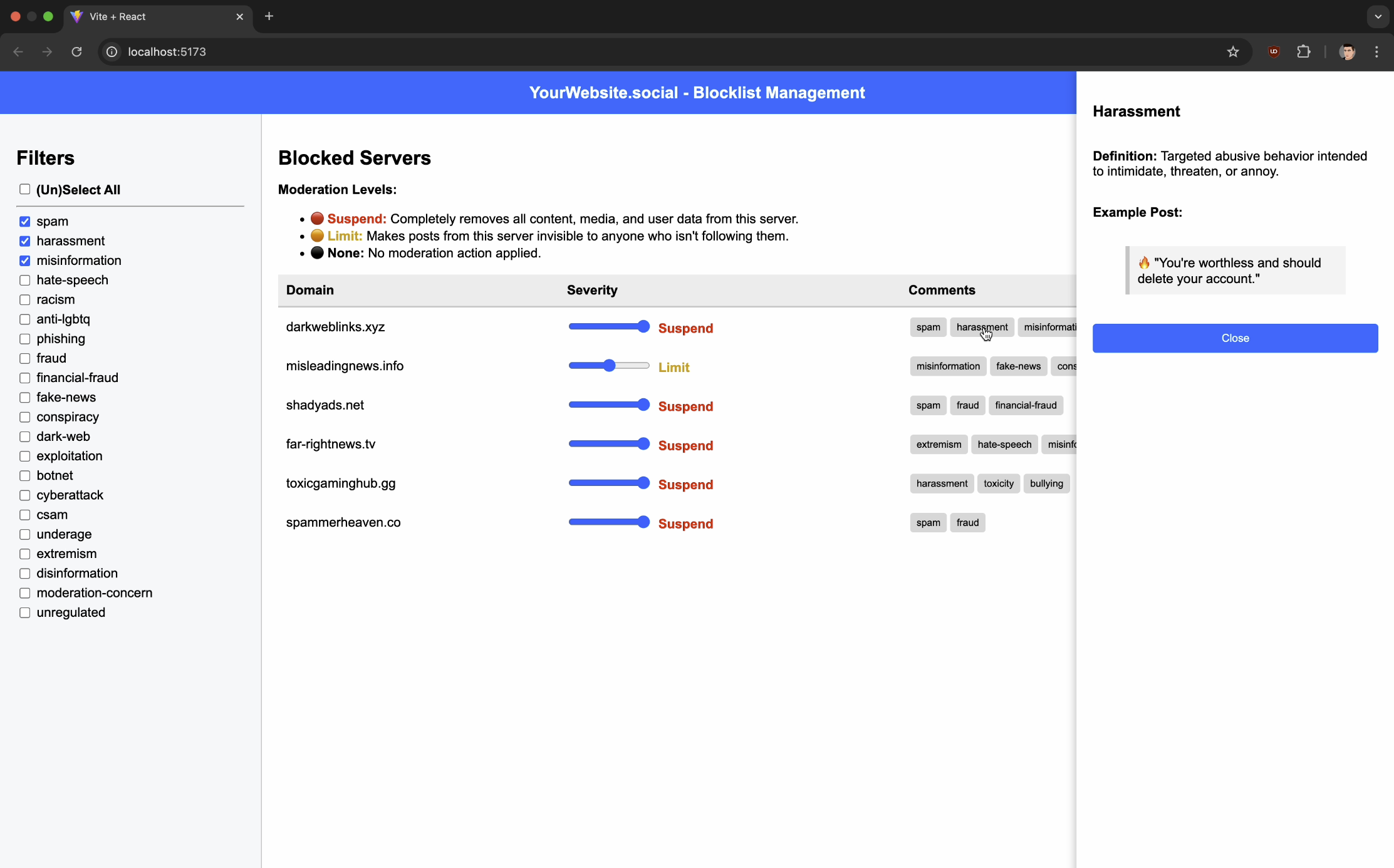}
    \caption{Screenshot of the demo system, showcasing category filters, severity toggles, and comments.}
    \label{fig:demo-system}
\end{figure}

\subsubsection{Demo system as probe.} 
\label{sub:demo_system}
We developed a lightweight web-based prototype system to support our exploration of improving blocklist management practices (Figure~\ref{fig:demo-system} shows the user interface). Built with React and Vite and styled using plain CSS, the system simulates moderation functionalities tailored for Mastodon moderators through three main features:

\begin{enumerate}
    \item \textbf{Category filters} enable moderators to selectively view instances based on specific comments, such as spam, harassment, or misinformation, displayed on the left-hand side panel.
    \item \textbf{Severity toggles} allow moderators to assign different action levels to each instance --- Suspend, Limit, or None --- each accompanied by an explanation. These actions are visually represented with red, yellow, and black indicators.
    \item \textbf{Comments} allow moderators to check the reason for blocking each instance. When users click on a comment tag (such as the cursor shown in the image), the system displays its definition and an example post in the right-hand side panel.
\end{enumerate}

\subsubsection{Data Collection and Analysis.}
We collected data from audio recordings of the interviews and survey responses. We transcribed interview recordings and stored them securely, with personally identifiable information redacted to ensure participant confidentiality. The first author independently conducted a thematic analysis \citep{braun2006using}, developing an initial codebook based on the research questions and exploratory goals. After coding all transcripts, the first author met regularly with the co-authors to iteratively refine the codes, review emerging themes, and ensure analytic consistency. Findings and interpretations were discussed among the authors, with iterative feedback cycles used to validate and refine the themes until reaching an agreement.

\section{Results}

We first present findings from our content analysis of \textit{community-level blocklists}, \textit{shared community-level blocklists}, and related \textit{blocklist-related tools} used by Mastodon communities, noting patterns and limitations (\S \ref{sec:findings_rq1}). We report findings from our interviews, which first describe current practices and perceptions around blocklists (\S \ref{sec:findings_rq2}) and then identify features that would facilitate improved use of blocklists (\S \ref{sec:findings_rq3}).

\subsection{The Current Landscape of Community-Level Blocklists (RQ1)}
\label{sec:findings_rq1}

Our content analysis suggests that \textit{community-level blocklists} are heterogeneous and often not publicly shared.
At the same time, those \textit{shared community-level blocklists} (Table \ref{tab:blocklists-summary}) aim to be ``starter'' resources that serve as foundations but vary widely in how they envision being useful to this end. They are limited in transparency and focus on English-language contexts. 
\textit{Blocklist-related tools} (Table \ref{tab:tools-summary}) offer technical facilitation --- usually to help people keep blocklists updated --- but often lack standardized mechanisms for community feedback and content categorization.

\subsubsection{Overview of Community-Level Blocklists.}

Out of approximately 8,700 Mastodon instances\footnote{Instance list from the organization that manages the Mastodon software available at \url{https://joinmastodon.org/servers}}, we analyzed 1,807 instances with at least 10 active users. Among these instances, only 364 (20.1\%) publicly shared their blocklists, suggesting that explicit transparency in moderation remains limited across Mastodon. Within this subset of 364 instances, fewer than half (169, or 46.4\%) provided explicit reasons for blocking. This finding indicates a gap in accountability and interpretability of moderation actions.

Table~\ref{tab:blocklist_by_size} summarizes the distribution of public blocklist sharing by instance size, illustrating a clear trend: larger instances were notably more likely to share their blocklists publicly. Specifically, the largest instances (1,000+ users) shared their blocklists at the highest rate (45.3\%), while the smallest ones (10–24 users) had the lowest sharing rate (13.9\%). This pattern suggests that transparency in moderation decisions increases with community size, possibly due to greater moderation resources or heightened expectations from larger user bases.

Table~\ref{tab:blocking_categories_pct} lists the top 10 blocking categories, derived from a thematic analysis of explicit reasons provided by moderators. Among the 169 instances that provided rationales, spam (69.8\%), harassment/trolling (50.3\%), and automated bots (47.3\%) emerged as dominant concerns, reflecting shared moderation priorities aimed at reducing disruptive and unwanted content. These results highlight consistent moderation challenges faced across Mastodon communities.

\subsubsection{Shared Community-Level Blocklists.}

The blocklists analyzed share an emphasis on manual curation and draw from external sources, though these are often not publicly disclosed. Their scopes range from consensus-based inclusion to severity-focused targeting. Distribution methods vary, including synchronization tools and public CSVs. Common limitations include subjective criteria, regional and language biases, and infrequent updates due to the labor-intensive nature of review.

\xhdr{Purpose and Criteria.} The blocklists analyzed typically serve as foundational ("starter") moderation resources but vary significantly in their strategic focus, ranging from broad consensus to high-severity domains. General-purpose blocklists such as \textit{Garden Fence} and \textit{CARIAD} include widely acknowledged harmful domains sourced from multiple moderation communities, prioritizing broad protection. In contrast, \textit{IFTAS-DNI} and \textit{FediNuke} explicitly target domains deemed severely harmful or notoriously problematic. Notably, selection criteria across all blocklists rely on external Mastodon instances for decisions rather than solely on internal judgments. However, these external sources and their blocklists are typically not publicly shared. For example, \textit{Seirdy Tier-0} explicitly adopts a consensus-based approach (requiring domains to appear on at least 15 out of 27 independent blocklists), while withholding the identities of the source lists.

\begin{table}[]
  \centering
  \footnotesize
  \caption{Public Blocklist Sharing and Top Blocking Categories}
  \label{tab:side_by_side}
  \begin{subtable}[t]{0.48\textwidth}
    \centering
    \caption{Public Sharing by Instance Size}
    \begin{tabular}{lrrr}
      \toprule
      \textbf{Instance Size} & \textbf{Yes} & \textbf{No} & \textbf{\%} \\
      \midrule
      10--24   &  89 & 549 & 13.9\% \\
      25--49   &  62 & 288 & 17.7\% \\
      50--99   &  51 & 188 & 21.3\% \\
      100--199 &  50 & 148 & 25.3\% \\
      200--499 &  42 & 120 & 25.9\% \\
      500--999 &  20 &  29 & 40.8\% \\
      1000+    &  34 &  41 & 45.3\% \\
      \bottomrule
    \end{tabular}
    \label{tab:blocklist_by_size}
  \end{subtable}
  \hfill
  \begin{subtable}[t]{0.48\textwidth}
    \centering
    \caption{Top 10 Blocking Categories}
    \label{tab:blocking_categories_pct}
    \begin{tabular}{lrr}
      \toprule
      \textbf{Category}          & \textbf{Count} & \textbf{\% of 169} \\
      \midrule
      Spam                       & 118 & 69.8\% \\
      Harassment / Troll         &  85 & 50.3\% \\
      Bots                       &  80 & 47.3\% \\
      Hate Speech                &  64 & 37.9\% \\
      CSAM / Child Abuse         &  57 & 33.7\% \\
      Misinformation             &  50 & 29.6\% \\
      Facebook/Meta              &  40 & 23.7\% \\
      Transphobia                &  37 & 21.9\% \\
      Adult / NSFW               &  37 & 21.9\% \\
      Racism                     &  30 & 17.8\% \\
      \bottomrule
    \end{tabular}
  \end{subtable}
\end{table}

\begin{table}[]
\centering
\vspace{2mm}
\caption{Summary of Mastodon Blocklists}
\label{tab:blocklists-summary}
\scriptsize
 \begin{tabularx}{\textwidth}
{|>{\raggedright\arraybackslash}p{1.5cm}|>{\raggedright\arraybackslash}X|>{\raggedright\arraybackslash}X|>{\raggedright\arraybackslash}X|>{\raggedright\arraybackslash}X|>{\raggedright\arraybackslash}X|}
\hline
\textbf{Category} & \textbf{Garden Fence} & \textbf{CARIAD} & \textbf{IFTAS-DNI (Do Not Interact)} & \textbf{Seirdy Tier-0} & \textbf{FediNuke} \\
\hline
Purpose & Starter blocklist protecting users from well-known harmful sources. & Initial moderation resource for widely-blocked domains. & Curated subset of CARIAD domains recommended for defederation. & Customizable starter blocklist aligned with Mastodon Covenant standards. & Recommended default blocklist focusing on severe and notorious instances. \\
\hline
Criteria & Includes domains blocked by \texttt{sunny.garden} and other instances. (not publicly shared) & Combines data from the IFTAS-DNI list and major Mastodon providers. & Labels domains according to well-established moderation glossaries \cite{dtsp2023glossary}. & Consensus-based inclusion (appearing on $\leq$ 15 of 27 source blocklists). & Severity-based subset, prioritizing severe, notorious actors. \\
\hline
Transparency & Domains reviewed manually, reference instances not publicly disclosed. & Includes domains with 51\% consensus, manually reviewed by IFTAS. & N/A & Curated by accountable individual, sources selected privately with overrides. & Manually curated, severity-focused, supported by strict receipt criteria. \\
\hline
Distribution & Publicly available, updated weekly. & Available via moderation tools such as FediCheck. & Publicly accessible online via the moderators' forum. & Public CSV form, periodically updated. & Public CSV form, updated less frequently and selectively. \\
\hline
Limitations & Subjective judgments, English-language bias. & Biased toward global north perspectives; lacks marginalized community protection. & N/A & Subjective, limited transparency, slow manual updates. & Inherits Tier-0 limitations, intentionally limited completeness and updates. \\
\hline
Other Resources & Recommends additional blocklists and moderation tools. & Suggests tools for synchronizing Mastodon blocks (e.g., FediBlockHole). & N/A & Suggests FediBlockHole, supplementary blocklists. & Same as Tier-0. \\
\hline
Additional Content & N/A & Public feedback channels provided. & N/A & Documents moderation evidence ("receipts"), manual updates, community feedback. & Same as Tier-0. \\
\hline
\end{tabularx}

\end{table}
\vspace{50em}

\xhdr{Transparency.} Even for widely shared blocklists, details such as the specific instances they rely on are often withheld. Lists such as \textit{Seirdy Tier-0} and \textit{Garden Fence} offer relatively detailed documentation, including moderation rationales ("receipts") and criteria for overrides. In contrast, blocklists like \textit{IFTAS-DNI} do not disclose their internal decision-making processes, limiting transparency. Furthermore, \textit{Garden Fence} explicitly incorporates external blocklists that are not publicly shared, contributing to opacity despite efforts to communicate manual review procedures.

\xhdr{Distribution and Additional Tools.} Most blocklists are updated regularly, and some are integrated into moderation workflows. Blocklists such as \textit{Garden Fence} and \textit{CARIAD} are distributed via synchronization tools like FediCheck and FediBlockHole, enabling convenient and rapid adoption. Others, such as \textit{Seirdy Tier-0} and \textit{FediNuke}, are distributed as publicly accessible CSV files, supporting more manual but transparent integration.

\xhdr{Limitations.} All analyzed blocklists exhibit inherent limitations tied to manual curation practices, including subjectivity, linguistic and cultural biases, and challenges with timely updates. For instance, \textit{Garden Fence} openly acknowledges biases toward English-language sources and subjective administrative decisions. Similarly, \textit{CARIAD} notes biases toward Global North perspectives, potentially disadvantaging marginalized communities. \textit{Seirdy Tier-0} emphasizes the intensive labor involved in manual curation, acknowledging delays in updates and responsiveness. Collectively, these limitations highlight moderation trade-offs between accuracy, inclusivity, and timeliness, raising important considerations for communities adopting blocklists.

\subsubsection{Blocklist-Related Tools.}

The tools analyzed support the discovery, evaluation, and use of blocklists, streamlining moderation through automation, curation, and categorization. They all aim to simplify blocklist management but differ in scope, access, and decision-making structure.

\xhdr{Purpose.}
All tools support community moderation, but their primary functions differ. \textit{FediCheck} and \textit{FediBlockHole} offer “moderation-as-a-service” by allowing moderators to sync with trusted blocklists quickly. In contrast, \textit{Fediseer} and \textit{The Bad Space} emphasize classification and transparency. They provide curated or crowd-sourced records of harmful or under-moderated instances, serving as reference directories for moderators.

\xhdr{Technical Access.}
The tools offer different modes of access, including web dashboards and APIs. \textit{FediCheck} and \textit{The Bad Space} provide web interfaces. \textit{FediBlockHole} operates as a script or service using the Mastodon moderator API and allows importing and exporting blocklists from URLs, files, or other instances. \textit{Fediseer} offers both a web interface and a machine-readable API, enabling direct use and integration with other moderation tools.

\xhdr{Decision Type.}
Tools also vary in how they structure decision-making authority. \textit{FediCheck} automates the application of IFTAS-curated lists (CARIAD and DNI) but leaves the final choice to the user. \textit{FediBlockHole} lets moderators subscribe to blocklists from other instances while retaining override control. \textit{Fediseer} tracks community-based judgments but does not implement blocks. \textit{The Bad Space} lists only those instances flagged by at least two independent moderation communities, reducing reliance on single-source decisions.

\begin{table}[t]
\renewcommand{\arraystretch}{1.3} % optional: adds line spacing
\centering
\caption{Summary of Mastodon Blocklist-related Tools}
\footnotesize
\begin{tabularx}{\textwidth}{|>{\raggedright\arraybackslash}p{1.5cm}|
                                >{\raggedright\arraybackslash}X|
                                >{\raggedright\arraybackslash}X|
                                >{\raggedright\arraybackslash}X|
                                >{\raggedright\arraybackslash}X|}
\hline
\textbf{Category} & \textbf{FediCheck} & \textbf{Fediseer} & \textbf{FediBlockHole} & \textbf{The Bad Space} \\
\hline
Summary & 
Moderation-as-a-service, automatically syncing blocklists for Mastodon moderators. &
Moderation tool using a crowdsourced endorse/censure system to flag spam and other harmful behavior. &
Open-source tool auto-syncing Mastodon instance blocklists from trusted sources. &
Community-driven catalog identifying harmful instances via marginalized communities’ consensus. \\
\hline
Purpose &
Helps new admins quickly implement moderation by auto-applying vetted denylist. &
Supports community-based moderation through a “chain of trust” for spam and other behaviors. &
Simplifies moderation by aggregating trusted blocklists. &
Lists instances posing risks due to poor moderation or harmful content. \\
\hline
Technical Access &
Web app (Mastodon v4.1+ API) for remote domain block management. &
Web interface and machine-readable API. &
Service/script via Mastodon Moderator API for fetching/pushing blocklists. &
Web-based searchable directory, public API for flagged instance data. \\
\hline
Decision Type &
Automates blocks using aggregated and moderated lists (CARIAD/DNI). &
Uses three judgment types: guarantees, endorsements, censures. &
Allows subscriptions to external lists with moderator final control. &
Lists instances only after independent decisions by multiple communities. \\
\hline
Tags / Categories &
Uses standardized DTSP labels for classifying behaviors and content. &
Allows up to 100 descriptive tags per instance for filtering/discovery. &
Does not provide own tagging; supports public/private comments. &
Categorizes reasons by harmful behavior types identified by communities. \\
\hline
\end{tabularx}
\vspace{8mm}
\label{tab:tools-summary}
\end{table}

\xhdr{Tags and Categories.}
Only some tools support structured tagging. \textit{FediCheck} uses standardized labels from the Digital Trust \& Safety Partnership (DTSP) Glossary \cite{dtsp2023glossary}, which categorizes harmful behavior (e.g., harassment, misinformation) with definitions and example use cases. \textit{Fediseer} allows instance moderators to apply up to 100 descriptive tags per entry to aid filtering and discovery. \textit{The Bad Space} uses clearly labeled categories focused on user safety concerns such as racism, transphobia, and ableism---defined in collaboration with its partner communities. \textit{FediBlockHole} does not implement formal tagging but allows optional comments to provide context.

\subsection{Moderator Practices and Perceptions of Blocklists (RQ2)}
\label{sec:findings_rq2}
We asked community moderators about how they made everyday decisions about blocking instances.
Moderators first identified potentially problematic instances to block in three key ways: 

\begin{enumerate}
    \item Moderators looked at publicly-shared \textbf{blocklists} and \textbf{related blocklist-related tools} such as those ones reviewed in \textbf{RQ1}. Among the five blocklists and four tools examined, P1, P2, P6, and P11 explicitly mentioned their use of the blocklist-related tool \textit{Fedicheck} (which references blocklists \textit{CARIAD} and \textit{IFTAS-DNI}); P9 used Seirdy's blocklist and The Bad Space as references.%, thinking of them as ``a kind of instance catalog.'' 
    \item Moderators often followed the \textbf{\#Fediblock hashtag}, a hashtag created by Artist Marcia X for people to collaboratively and publicly report instances that they have had negative experiences with \citep{logic2023blackness}; by following or routinely checking the hashtag, users can see potential instances they may want to consider blocking. ``'From time to time, I also check the \#FediBlock hashtag to see if there’s anything new I should be aware of''. (P9) Multiple interview participants (P1, P5, P6, P9) mentioned they follow the \#FediBlock hashtag. 
    \item Moderators relied on \textbf{user reports} to help them surface patterns that might serve as a ``kind of clue that, {Oh, maybe this server is hosting a lot of spam bots and stuff}'' (P1) that would prompt a moderator to investigate further. 
\end{enumerate}

Moderators noted a few strategies for investigating instances further. P1 said that sometimes just checked the instance's About page and public feed to get the ``\textit{vibe},'' looking for spammy, toxic, or otherwise misaligned content\footnote{e.g., NSFW content would not work for a professionally-oriented instance}. P10 focused on instances' \textit{rules}, checking ``if the rules [...] are specially catered to a group of right-wing fascists, or do not regulate hate speech at all'' (both of which would make an instance an immediate block candidate). Meanwhile, P11 emphasized considering \textit{admin behavior}: ``If we get no management (response) from the  [about a problem]—or the instance moderators themselves is the problem—those ones are the ones that we block.'' 
Several participants (P3, P5, P6, P9, P11, P12) said it was an easy choice to suspend instances that were clearly poorly-managed, e.g., full of spam, or allowed behavior ``clearly outside of the bounds of what [is] acceptable'' (P7) such as CSAM, spam, racism, harassment, and anti-trans behavior. 

However, not all decisions were straightforward. For example, a specific user's behavior could create what P7 called a ``borderline case'': serious enough transgressions to suspend the specific user (e.g., being uncivil or accidentally sharing misinformation), but not to block the whole instance as the user didn't ``step clearly outside of rules of the server.'' Additionally, moderators sometimes wanted to block instances because of scale and ownership rather than obvious content or behavior issues. For example, P1 and P5 wanted to block large instances like \texttt{mastodon.social} because they felt it had poor moderation practices, but blocking it meant cutting their community off from a huge number of users on that instance. 
How moderators navigated such nuances around community-level blocking varied. Interviews pointed to three types of decisions which reflected competing priorities shaping blocklist perception and use, elaborated below.

\subsubsection{Responsive, report-based decisions prioritizing openness.}
Some moderators focused on prioritizing openness, emphasizing the value of exposure on a decentralized social network in creating a richer community: ``Just like in New York, where you walk outside and you're exposed to everything... I try to embody that culture a little bit in the server'' (P8). This echoed broader sentiments about the purpose of a social network to begin with: connecting people. 
Interview participants voiced concerns that over-blocking could hurt new instances just learning how to moderate, as well as the reach of users from underrepresented communities, whose needs and contexts often become marginalized online, in particular: 
\begin{quote} 
``%I'm very sensitive to over-moderation ... 
One big issue is that people will brigade accounts that are doing nothing wrong, simply because they don’t like them. This is how many queer people and others from marginalized communities end up getting banned on social media: right-wing users mass report the account, and moderation systems typically ban it first. ... I’d worry about something similar happening here.'' (P4)
\end{quote}

\noindent Participants (P3, P4) who took this view also suggested that the blocklists made by others (including the collaborative resources we reviewed prior) were not always a good reference for their own community, because the values and judgment guiding them were different. 

Moderators prioritizing openness tended to take a more reactive approach, adding instances to their blocklist as needed. They thus relied heavily (but not exclusively) on user reports, and multiple participants tried to encourage active reporting among their community members. P11 told their users: ``If you’re not sure if you should report it, report it. But I would rather you show it to me than think that I wouldn’t do anything or think that I can’t do anything.'' P8 also sometimes changed blocking decisions based on community member feedback: ``At the end of the day, we only have two or three hundred active users. We [...] really value the direction they want to go.'' 

\subsubsection{Proactive, search-based decisions prioritizing safety.}
Moderators noted that ensuring community safety required some proactive effort to block potential bad actors, as there wasn't a built-in, ``systematic way to detect and react to servers'' (P10) hosting hate speech and other types of content one sought to avoid. Even if this led to over-blocking, some participants felt it was safer: 
\begin{quote}
    If I block an instance by mistake, it’s not the end of the world [... People] can reach out and let me know, or if someone on one of my instances asks, `Hey, why are you blocking them? I’ve got some friends there,' then I can go look into it ... At most, there’s some temporary loss of communication. To me, that’s far less risky than not blocking something that could lead to harassment.'' (P9)
\end{quote}

A handful of our participants also pointed out that proactive efforts, especially in ``an early stage of becoming a moderator'' (P4), could reduce time and labor later on. P5 stated: ``If I didn’t block these instances, then a few months later, their users would harass our users. And then we’d get a lot of reports, and I’d have to block them anyway.'' 
P8 --- who strongly valued openness in their instance --- also acknowledged the need to prevent burnout, particularly with new moderators: ``Feel free to start too restrictive and work your way back to your comfort level, because burnout is real [...] Moderating can feel like boiling the ocean—it doesn't have to be that way [...] The most important resource in a Trust and Safety role is yourself: your own energy, your mental health.''

Shared blocklists and the \#Fediblock hashtag appeared to be valuable tools to this end. P10 noted that they helped people anticipate harm and harassment, becoming a useful starting point for people just beginning to run an instance. Third-party tools (e.g., Table \ref{tab:tools-summary}) that would automatically check shared blocklists and then sync their own could eliminate the need to manually do so. P1 felt that Fedicheck made moderation much easier for their instance by automating blocklist updates: ``for a small server like ours, [we] don't have time.'' 

\subsubsection{Measured, manual-review decisions prioritizing context.}
The third type of decision-making emphasized making careful, manual reviews of what should and should not go in one's blocklist. 
P12 stressed that blocking instances impacted the entire community's connectivity, thus implying a ``serious responsibility'' of a moderator. P3 wanted to avoid ``[using] someone else's moral judgment to apply a block on an entire community of people,'' especially as they felt reasons for blocking can be arbitrary (e.g., use of a different language). %``any number of criteria that have very little to do with the actual purpose of the blocking''. 
P9 further noted that some of the best-known shared blocklists were not designed to ensure the safety of vulnerable communities: 
\begin{quote}
    ``IFTAS put together their CARIAD blocklist [which] was originally based on aggregating blocks from the most-used servers. They specifically stated that this blocklist will not protect Black, Asian, Eastern, and trans communities, because that’s not the philosophy of the top servers, which are generally run by cis, white guys.''
\end{quote}

Despite their critical views of external blocklists, participants noted they still used them periodically. For example, P12 said they would consider using blocklists if ``some neo-Nazi instances are popping up like crazy [and] I can no longer moderate the instances hand-to-hand, by my own effort.'' But overall, participants like P3, P9, and P12 encouraged manually reviewing external blocklists to check for false positives and whether blocks made sense to them. P1 and P11 annotated their blocklists with private and public comments per block. Private comments, or internal notes, are only visible to moderators, helping to explain to others—and to their future selves—the detailed reason behind each decision, making it ``justifiable, renewable, and reviewable.'' (P11) Public comments are viewable statements justifying blocks in relation to instance rules --- although often not made because it requires more effort.
Several participants indicated that a manual approach could require far too much work in general. P7 noted that tooling for blocklists on the Fediverse is still very limited, which is why they dislike using blocklists. 

\subsection{Moderator Needs to Facilitate Improved Blocklist Use (RQ3)}
\label{sec:findings_rq3}

Interviews with moderators about their use and view of blocklists pointed to limitations and challenges they needed to navigate. Thus, we also asked moderators what would improve blocklists, leveraging the demo tool we built as a design probe. Interviews pointed to three needs that would make blocklists more effective and powerful community-level tools.

\subsubsection{Community collaboration. }
Participants emphasized the importance of collective approaches to blocklist management, including voting on instances, as well as sharing, syncing, selecting, and subscribing to shared blocklists.  

Collaboration \textit{within} a community --- i.e., discussing the blocklist with members of the instance --- was a common focus. P1 proposed a voting system to enhance internal moderator communication and decision-making: ``Maybe you want to ask the team of moderators for their advice. And maybe we could have something like a voting system or a button to tell the team, 'We have a question on this one. Can we discuss? Or maybe chat, or leave notes.''' 
Managing multiple instances, P5 expressed interest in the ability to share blocklists and automate blocking decisions across instances they moderate, ``not publicly, just between our own instances.''  

Participants also discussed collaboration across communities, such as the value of having and leveraging collective signals. P5 proposed a public ``suspect list'' where moderators could comment and vote on instances' status. Similarly, P9 proposed a shared public receipt library similar to Seirdy's receipts and The Bad Space. 
P8 envisioned ``scorecards'' indicating how restrictive or permissive a blocklist is: ``for some servers who are very restrictive, `we have the safest block list'—you can see it here. [...] What if you were to go to a server and it said, here's our block list scorecard—we block 100\% of CSAM servers. That’s a strong statement that would appeal to almost everybody.'' 
P8 also suggested a ``marketplace of blocklists,'' where users could choose and subscribe to different blocklists tailored to their preferences.

\subsubsection{Efficiency and management. }
Participants highlighted the need for tools that enhance efficiency and streamline moderation workflows. 
In our demo system, many participants (P1, P5, P6, P8, P9, P11, P12) found the category filters useful. Participants thought the filters allowed moderators to “quickly spot problem content like spam and harassment” (P11), and described filters as taking “a lot of work off” (P5) their moderation workload and helping “spot obvious cases” (P12) without needing to investigate each instance manually. Filters also helped moderators organize their blocklists, such as by “cleaning up the list and making sure we’re focusing on real issues” (P6) or “categorizing servers into types” (P8). 

P1 emphasized that filtering addressed “something quite difficult to do right now—there’s no tools to do so,” highlighting the lack of built-in support in current moderation systems.
As a small instance owner, P1 also emphasized the need for automated services capable of ``finding every problematic server on the Fediverse,'' ideally integrated into the default Mastodon interface. P4 echoed this, saying, ``I'm sure something like this --- if you could click on something and it automatically blocks it from your instance --- would be helpful.''  

\subsubsection{Monitoring and visibility. } 
Participants discussed the need for richer, large-scale information that would support moderation decisions by helping them better monitor and assess cases. 
P1 emphasized several key metrics, such as instance activity levels, user population, and the number of their own instance members who follow the targeted instance. These metrics significantly influenced blocking decisions—for instance, P1 preferred not to block inactive instances, took larger instances more seriously, and considered disruptions caused by users losing followers. Similarly, P9 highlighted the importance of understanding whether other instances also blocked a given instance, reflecting how tools like Fediseer aggregate and show which instances have blocked a given instance. The number of user reports was also a critical signal for moderation decisions (P2). Additionally, P4 proposed using instance-level descriptive statistics to support moderators' judgments:

\begin{quote}
``I think it would be helpful if there were a way to examine large-scale, instance-level trends—maybe something averaged across users—to get a better sense of what’s going on at the domain level. [...] I’d want to know how many people are on a given domain, their post or comment frequency, and maybe some kind of quantitative measure that gives a high-level sense of the content.'' (P4)
\end{quote}

Finally, participants responded positively to comment features on our demo and suggested improvements that would enhance existing moderation documentation tools. 
P2, P5, P6, and P8 found the comment feature particularly valuable, highlighting its role in enhancing transparency for both moderators and community members by providing clear documentation (P2, P5). Participants also used comments to organize instance actions, like ``categorizing servers into types'' (P8), and to help moderators “remember why we blocked that instance” (P6) or revisit decisions ``based on the content of the report'' (P2).

Several participants also recommended further enhancements. P10 recommended multilingual support for custom tags to accommodate non-English-speaking communities, and others emphasized the need for a detailed ``receipt'' system, including blocking dates, sources, and notes for traceability—so moderators could track “who did what, and when” (P1), understand “if a block was inherited” (P7), or maintain a shared public “receipt library” (P9) for moderation decisions.

\section{Discussion}
Our findings (\textbf{RQ3}) describe extant practices and perceptions around blocklists, considering both how communities add to their blocklists and the public external resources they can leverage. Our interviews (\textbf{RQ2}) noted how communities balance competing priorities around openness, safety, and context in making moderation decisions with community-level blocks, which can be much more impactful than individual-level blocks. Moreover, combined with our content analysis (\textbf{RQ1}), we observed some current limitations that make blocklists difficult, laborious, or time-consuming to use meaningfully and well. Finally, we surfaced avenues for future design work that would improve the use of blocklists. Here, we discuss the broader implications of our findings for moderation practices with community tools.

\subsection{Balancing Competing Priorities in Moderation}
\begin{figure}[t]
    \centering
    \begin{tikzpicture}[font=\small, align=center]

        \node (Context) at (0,2.5) {\textbf{Context}\\
        Nuanced fairness\\
        \textit{Risk: labor-intensive}};

        \node (Openness) at (-4,-1.5) {\textbf{Openness}\\
        Connection, free expression\\
        \textit{Risk: harmful exposure}};

        \node (Safety) at (4,-1.5) {\textbf{Safety}\\
        Harm prevention\\
        \textit{Risk: penalizing good actors}};

        \draw[<->, thick] (Openness) -- (Safety) 
            node[midway, below, yshift=-0.3cm] {Balance of free expression\\vs. harm prevention};

        \draw[<->, thick] (Safety) -- (Context) 
            node[midway, right, xshift=0.3cm] {Balance of efficiency\\vs. fairness};

        \draw[<->, thick] (Context) -- (Openness) 
            node[midway, left, xshift=-0.3cm] {Balance of nuanced decisions\\vs. exposure risks};

    \end{tikzpicture}

    \caption{Conceptual diagram showing the triangular balance among moderation priorities: \textit{Context}, \textit{Openness}, and \textit{Safety}. Each approach comes with inherent risks, and moderation requires carefully balancing these competing priorities.}
    \label{fig:triangular_balance}
\end{figure}
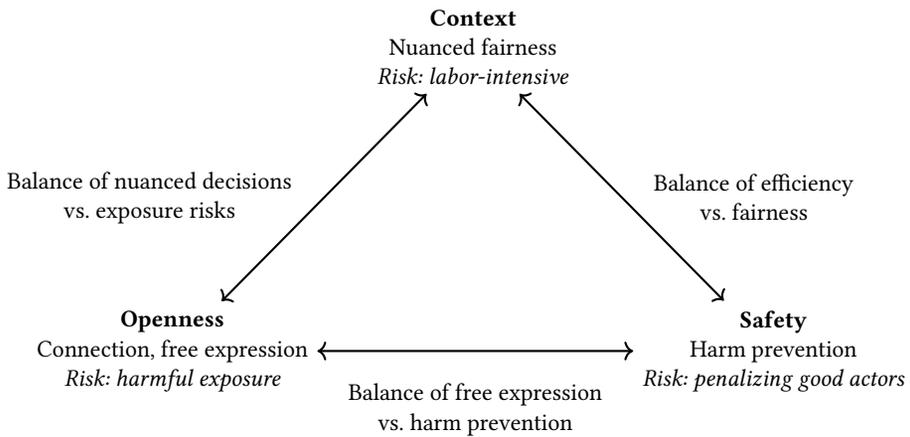

We noted three approaches to decision-making around blocking that reflected different priorities: \textit{openness}, \textit{safety}, and \textit{context}. Our interviews pointed to benefits and drawbacks of each: openness enabled connection, but could risk exposure to bad actors; safety helped ensure positive online interactions, but could penalize good-faith actors; context could improve the fairness of blocking decisions, but could be incredibly labor-intensive. We illustrated them in Figure~\ref{fig:triangular_balance}.
These insights resonate with prior work evaluating proactive vs. reactive moderation strategies \citep{Schluger2022Proactive,Jiang2023Tradeoffs,seering2019moderator,seering2022metaphors}, particularly work noting that achieving both openness and safety at once has been a key challenge in moderation on social media. For example, Jiang and colleagues~\citep{Jiang2023Tradeoffs} highlight how content moderation research often takes either ``nurturing'' and ``punitive'' philosophies, reflecting a deeper normative conflict between proactively preventing harm and protecting free expression. 

However, scholars also note that proactive strategies focused on safety and reactive strategies focused on openness can and should be used in tandem \cite{kiesler2012regulating, Schluger2022Proactive}. Likewise, our findings suggest that \textbf{the three approaches to decision-making prioritizing openness, safety, and context --- although competing --- are not mutually exclusive}: participants appeared to engage in or switch between approaches, making both report-based decisions \textit{and} search-based decisions, along with manual-review decisions. Moreover, across the three approaches to decision-making, participants shared concerns that suggest they are not incompatible in goals or values. For example, participants were consistently concerned about how blocking as a community tool could protect or inadvertently punish marginalized and vulnerable communities, even in the reactive approach prioritizing safety --- which contrasts with prior work which often notes the limitations of reactive moderation strategies in ensuring safety \citep{jhaver2018blocklists, Chandrasekharan2019Crossmod, Schluger2022Proactive, kiesler2012regulating, blackwell2018online, dosono2019moderation, wohn2019volunteer}. 
Our findings underscore the importance of combining these three moderation approaches to offset each approach's weaknesses and leverage their complementary strengths. Rather than viewing openness, safety, and context as competing goals, communities can benefit from a hybrid approach that strategically integrates reactive openness-oriented moderation, proactive safety measures, and context-driven manual reviews. Ultimately, how communities balance these approaches will depend on their specific values, goals, and resources, highlighting the need for moderation tools that support flexible, context-sensitive configurations.

\subsection{Enhanced Responsibility Around Community-level Blocks}

Because community-level blocks could be far more consequential than blocking a single individual, our findings emphasized \textbf{an awareness for moderators to be able to manage blocklists more responsibly}. Prior work has described the roles moderators play on centralized social media --- such as \textit{guardians}, \textit{gatekeepers}, and \textit{janitors} \citep{seering2022metaphors} --- noting the significant power moderators can wield \citep{matias2019civic,matias2016goingdark, schneider2022implicit}. Within decentralized platforms, the role and influence of moderators become even more critical and amplified due to each instance independently setting its own moderation policies. Blocking on the Fediverse can effectively isolate entire communities \citep{spencer2025labour}, which has been useful in collectively reducing the reach of communities identified as toxic or enabling hate speech \citep{colglazier2024effects, lai2025defederation}, but in turn also suggests that false positives can be consequential.

Our findings emphasize moderators' heightened awareness of their responsibilities regarding community-level blocks and reveal their deliberate efforts to engage community members as active stakeholders in governance decisions. 
While prior work highlights the scarcity or inadequacy of explicit governance conversations \citep{Schneider2021Modular, hwang2024adopting, frey2023effective}, we found moderators in decentralized communities aware of their responsibility and seeking robust dialogue with their communities. This suggests the key barrier is not moderators' awareness or willingness, but rather the inadequacy of current moderation tools in supporting meaningful, transparent, and collective decision-making. Future research should explore mechanisms and tools that support sustained dialogue and consensus-building within communities, enabling moderators to execute their roles with greater responsibility, transparency, and alignment with community values.

\subsection{Designing for Collaborative and Transparent Moderation with Blocklists}
Our content analysis (\S \ref{sec:findings_rq1}) and interviews (\S \ref{sec:findings_rq2}, \S \ref{sec:findings_rq3}) all pointed to a need to improve current tooling and support around blocklists for community moderation. %For example, while some moderators value external blocklists (Table \ref{tab:blocklists-summary}) for their efficiency, collective safety, and support for informed decision-making through shared resources, others express skepticism, raising concerns about biases, transparency, and the necessity of contextual decision-making. 
Here, we propose two broader directions for future design based on the challenges and limitations identified in our work.

\subsubsection{Supporting Collaborative Moderation Across Communities.}

Our findings showed \textbf{a significant demand for collaborative moderation features explicitly designed to facilitate inter-instance communication and cooperation}, despite the availability of several third-party moderation tools across Mastodon instances. Participants specifically emphasized the need for improved mechanisms to discover, compare, subscribe to, and merge external blocklists—envisioning a robust ``blocklist marketplace'' (P8). While similar ideas have been explored on platforms like Bluesky through their ``starter packs,'' \footnote{Bluesky starter packs: \url{https://bsky.social/about/blog/06-26-2024-starter-packs}.} Mastodon instances involve greater complexity than individual user profiles, complicating the succinct representation of an instance’s moderation profile or reputation.

To address these challenges, designers should enhance structured documentation of moderation actions, such as moderation receipts, to increase transparency and accountability. Effective moderation documentation should include richer contextual metadata such as the involved moderators (actors), timestamps (timing), and justifications for actions. To foster inter-community knowledge sharing \citep{Schneider2021Modular}, more adoptions like publicly shared receipt libraries (e.g., Seirdy’s Receipts or The Bad Space), universal vocabulary tags recognized across multiple instances, and multilingual support can be used to enhance inter-instance collaboration.

Additionally, decentralized platforms rely on shared norms and deliberate consensus-building processes to govern effectively. The concept of a ``digital covenant'' highlights the importance of establishing mutual agreements among community members to maintain and enforce moderation standards~\cite{zulli2023digital}. We recommend that designers build explicit support for consensus-building mechanisms into moderation tools, such as voting interfaces, deliberative forums, or reputation systems, to enable communities to collaboratively define and uphold these shared norms transparently and democratically.

\subsubsection{Enhancing Trustworthy Transparency in Moderation Systems.}

Our findings revealed that experienced moderators frequently hesitate to adopt or trust external blocklists due to insufficient contextual information about how these lists are created and maintained. Moderators expressed the need for richer, domain-level metrics—such as instance activity levels, user population size, moderation histories, number of received reports, and reputation scores—to support informed decision-making. They further indicated a desire for explicit metadata about blocklist sources, selection criteria, and potential biases or limitations to enable critical evaluation before adoption.

Based on these insights, we recommend that \textbf{moderation tools provide clear and detailed analytics to moderators, enhancing their capacity to make transparent and informed decisions}. Additionally, these tools should facilitate explicit documentation of moderation rationales, including selection criteria, reasons for blocking, and categories of blocked content. Clearly documented moderation decisions and publicly available blocklists would complement an instance's homepage and stated community rules, offering prospective users greater flexibility and valuable context when selecting instances to join.

These design recommendations align closely with prior literature, emphasizing transparency as crucial to moderation governance. Transparency has been shown to establish legitimacy through openly justified actions~\cite{gillespie2018custodians}, enhance user trust~\cite{binns2018reducing}, and mitigate perceptions of unfairness~\cite{epstein2022explanation}. Prior work has also proposed shared safety mechanisms, such as federated trust networks among moderators and transparent moderation data~\cite{roth2024federated}. Moreover, clear visual explanations of system-generated decisions significantly improve user understanding and trust~\cite{eslami2018algorithmic}, further underscoring the importance of detailed documentation in enhancing the trustworthiness and effectiveness of moderation practices.

\section{Limitations}
Our study has several limitations that future work should consider. The small sample size of 12 Mastodon moderators limits the generalizability of our findings across the broader moderator group (e.g., beyond Mastodon instances).
To supplement this, we conducted a content analysis before our interviews to offer a deeper overview of well-known resources across the Mastodon ecosystem. However, our focus on a curated selection of publicly visible and widely referenced blocklists and tools may introduce selection bias and overlook less prominent resources (e.g., non-English resources). 
Given the use of decentralized social media in non-English speaking communities with different cultural and legal standards, future work should consider how our findings translate. Additionally, the dynamic and evolving nature of moderation practices in decentralized networks makes it challenging to capture a complete and stable picture; specific practices and perceptions may shift over time. However, we believe the general patterns captured in our work --- such as the competing priorities and design needs around collaboration, efficiency, and visibility --- speak to broader, persistent problems that future work must grapple with. 

\section{Conclusion}

Practices such as community-level blocklists in Mastodon present new opportunities and challenges for community-level moderation. Through content analysis and interviews with Mastodon moderators, our study highlights how community-level blocklists serve as essential tools for community moderation while reflecting diverse moderation philosophies and practices. We found that moderators take and move between three approaches in using blocklists, aiming to balance priorities around openness, safety, and context. However, several challenges persist in using blocklists as community-level tools. We identify key opportunities for designing blocklist management tools that are flexible, transparent, and collaborative, toward supporting more resilient decentralized communities. As decentralized networks evolve, understanding how to balance openness, safety, and context—and recognizing the heightened responsibilities of moderators—will be essential to fostering effective, collaborative, and inclusive moderation.

\section{Disclosure of the usage of LLM}
We used ChatGPT (GPT4o model~\cite{achiam2023gpt}) to facilitate the writing of this manuscript. 
The usage included:
\begin{itemize}
    \item Turn CSV format tables into LaTeX format tables.
    \item Look for and correct grammar errors.
    \item Polish existing writing done by authors using prompts such as ``Find a synonym for X'', ``Add transition words to this sentence,'' and ``Shorten this sentence without altering its meaning.''
\end{itemize}
\noindent
We did not use ChatGPT to write original text in the paper at hand.

\begin{acks}
We sincerely thank Jon Pincus and Shagun Jhaver for their thoughtful feedback and valuable suggestions. We also deeply appreciate all our interview participants who generously shared their insights and experiences with us.
\end{acks}

\bibliographystyle{ACM-Reference-Format}
\bibliography{0_reference}

\appendix

\end{document}